\documentclass[12pt]{article}
\usepackage{amssymb}
\usepackage{epsfig}
\usepackage{graphicx}
\usepackage{amsmath}
\usepackage{verbatim}

\csname @addtoreset\endcsname{equation}{section}
\begin{document}
\begin{titlepage}
\title{\bf\Large  New Constraints on Higgs-Portal Scalar Dark Matter  \vspace{18pt}}

\author{\normalsize Huayong Han and Sibo Zheng \vspace{12pt}\\
{\it\small   Department of Physics, Chongqing University, Chongqing 401331, P.R. China}\\}

\date{}
\maketitle \voffset -.3in \vskip 1.cm \centerline{\bf Abstract}
\vskip .3cm
The simplest Higgs-portal dark matter model, 
in which a real scalar singlet is added to the standard model, 
has been comprehensively revisited,
by taking into account the constraints from 
perturbativity, electroweak vacuum stability in the early Universe,
dark matter direct detection, 
and Higgs invisible decay at the LHC.
We show that the {\it resonant mass region} is totally excluded 
and the {\it high mass region} is reduced to a narrow window 
$1.1$ ~TeV $\leq m_{s} \leq$ $ 2.55$~ TeV,
which is slightly reduced to $1.1$~TeV $\leq m_{s} \leq$ $ 2.0$~ TeV 
if the perturbativity is further imposed.
This {\it high mass region} can be fully detected by the Xenon1T experiment.
\vskip 5.cm \noindent October  2015
 \thispagestyle{empty}

\end{titlepage}
\newpage

\section{Introduction}

The Standard Model (SM) as the effective theory below the electroweak (EW) scale has been established after the discovery of SM Higgs scalar with mass around $125$ GeV \cite{1202.1408,1202.14888} at the Large Hadron Collider (LHC). 
Unfortunately, there is no viable candidate for a dark matter (DM) in the SM,
which implies that an extension beyond the SM is necessary.

Among viable extensions the simplest Higgs-portal dark matter (HDM) \cite{Zee,0702143,0106249,0003350,0011335} is the most economic and of special interest from both the DM- and LHC-phenomenology.
In this extension, only a real singlet scalar $s$ is added to the SM.
With the aid of $Z_2$ parity which is needed for the stability of singlet DM,
the numner of model parameters is reduced to three.
It implies that the minimal model is very predictable.
In the subsequent studies on the HDM \cite{0405097,0609081,0611014,0706.4311,0710.2416, 0811.0658, 0909.0520, 0912.5038,1006.2518,1009.5377,1102.3024,1112.3299,1205.3169,1306.4710},
the parameter space is analyzed in terms of experimental constraints as follows.
\begin{itemize}
    \item Dark matter relic density \cite{1303.5076}, 
which shows that the {\it low mass region}, {\it resonant mass region}, and {\it high mass region} are all viable.
    \item Dark matter direct detection such as Xenon100  \cite{Xenon100}, Xenon1T \cite{ Xenon1T} and LUX \cite{LUX} experiments, which are able to effectively detect the {\it high mass region} up to $\sim 3$ TeV.
    \item Limit on the Higgs invisible decay $h\rightarrow ss$, 
which can effectively detect the {\it low mass region} with $m_{s}$ of order a few GeV.
\item Indirect detection such as limit on $\gamma$-ray lines from the Fermi-LAT \cite{1305.5597} is rather efficient to constrain the {\it resonant mass region} \cite{1412.1105,1508.04418, 0810.4267}.
\end{itemize}

However, these constraints are not sufficient, 
as some possibly strong theoretic ones have been ignored.
The object of this paper is to take into account all known experimental and theoretic constraints. 
Simultaneously, experiments constraints will be updated, if available.
In compared with the previous works, e.g., a comprehensive review in \cite{1306.4710}, 
our main differences are the following ones.
\begin{itemize}
    \item The SM EW vacuum suffers from large quantum fluctuations in the early Universe as induced by the inflation,
which imposes {\it strong} constraint \cite{vacuum1, vacuum2, vacuum3} on the model parameters. 
SM vacuum stability must be taken into account to constrain the parameter space. 
For an earlier attempt, see \cite{1407.6015,1112.3647}.
 \item Assume the HDM as an effective theory between the EW and Plank scale, 
perturbative analysis should be considered seriously,
which leads to {\it strong} upper bounds on the quartic coupling constant of $s^{2}\mid H\mid^{2}$ and the DM self-coupling constant.
This effective theory can be applied, e.g., to S-inflation, 
where  the scalar singlet DM is identified as the inflaton \cite{1507.03600}. 
    \item Very recently, the ATLAS collaboration has reported the latest limit about the Higgs invisible decay width \cite{invdecay} in the low and resonant mass regions.
This will be updated in our analysis.
\end{itemize}

The paper is organized as follows.
In Sec.2, we introduce notation and conventions in the HDM. 
In this section the constraint on the model parameters from the requirement of perturbativity is studied.
In Sec.3, we discuss the constraint on the model parameters from the SM vacuum stability for 125 GeV Higgs mass.
No additional mechanism for stabilizing the SM vacuum stability is employed except the DM scalar.

In Sec.4 we re-examine the DM phenomenology.
In particular, we calculate the $s$ scalar relic density 
and the spin-independent nucleon-DM scattering cross section by using MicrOMEGAs \cite{1407.6129},
which are consistent with results in some earlier literature.
In Sec.5, we update the LHC phenomenology about scalar DM,
by focusing on the latest limit on the Higgs invisible decay.

Finally, we present our main conclusions in Sec.6.
We find that the {\it resonant mass region} is totally excluded 
and the {\it high mass region} is reduced to a narrow window 
$1.1$ ~TeV $\leq m_{s} \leq$ $ 2.4$~ TeV.
If the requirement of perturbativity is further imposed,
the allowed mass range is slightly reduced to $1.1$~TeV $\leq m_{s} \leq$ $ 2.0$~ TeV.
In either case the mass window can be fully detected by the Xenon 1T experiment.

\section{HDM within Perturbative Region}
The Lagrangian for the simplest HDM with a $Z_{2}$ parity,
under which $s$ is odd and SM particles are even, is given by,
\begin{eqnarray}{\label{Lagrangian1}}
\mathcal{L}=\mathcal{L}_{\text{SM}}+\frac{1}{2}\left(\partial s\right)^{2}+V(s, H)
\end{eqnarray}
where 
\begin{eqnarray}{\label{Lagrangian2}}
V(s, H)=\frac{\lambda}{2}\left(\mid H\mid^{2}-\frac{\upsilon^{2}}{2}\right)^{2}
+\frac{\mu^{2}_{s}}{2}s^{2}+\frac{\lambda_{s}}{2}s^{4}+\frac{\kappa_{s}}{2}s^{2}\mid H\mid^{2}.
\end{eqnarray}
In Eq.(\ref{Lagrangian2}) the first term denotes the Higgs potential with EW scale $\upsilon\simeq 246$ GeV, 
and the $\lambda_s$-term and $\kappa_s$-term refers to $s$ self-interaction 
and Higgs-DM interaction, respectively. 
Expand the fields as $H=(\upsilon +h)/\sqrt{2}$ and $s=\left<s\right>+s=0+s$ along the vacuum structure for positive $\lambda_s$ and $\kappa_s$
\footnote{The vacuum is given by $\left<H\right>=\upsilon /\sqrt{2}$ and $\left<s\right>=0$ for both positive $\lambda_s$ and $\kappa_s$,
which may be violated when both of them are negative.
However, as we will shown in Sec.3, 
neither negative $\kappa_s$ and nor negative $\lambda_s$ are favored by the EW vacuum stability. 
In this paper we choose both positive $\lambda_s$ and $\kappa_s$ for our discussion. }, 
one obtains from Eq.(\ref{Lagrangian2}), 
\begin{eqnarray}{\label{Lagrangian3}}
V(s, h)= \frac{1}{2}m_{s}^{2}s^{2}+\frac{\lambda_{s}}{2}s^{4}
+\frac{\kappa_{s}\upsilon}{2}s^{2}h+\frac{\kappa_{s}}{4}s^{2}h^{2}.
\end{eqnarray}
where $m_{s}=\mu_{s}^{2}+\kappa_{s}\upsilon^{2}/2$.
Remarkably, there are only three new parameters in the HDM.
It implies that this model can be studied concretely.

In some situations peturbativity should be imposed on the model.
This is true when it is assumed to the effective theory below some high energy scale such as Plank mass scale. 
See, e.g. \cite{1507.03600} for an application of such idea to the S-inflation.

\begin{figure}[!h]
\centering
\includegraphics[width=4.5in]{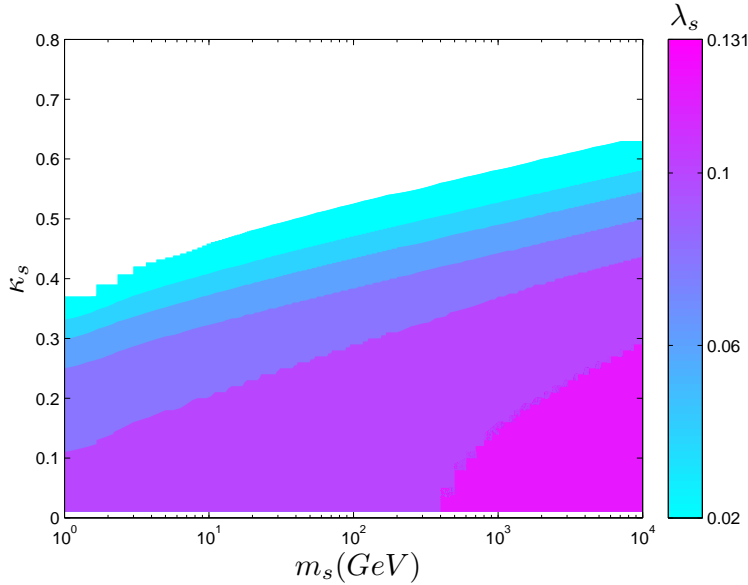}
 \caption{Upper bounds on the EW values of $\kappa_s$ and $\lambda_s$ from  the requirement of perturbativity below the Plank scale.}
 \label{figp}
\end{figure}

In Fig.\ref{figp} we show the upper bounds on the EW values of $\kappa_{s}$ and $\lambda_s$ by
the requirement of perturbativity
\begin{eqnarray}{\label{perturbativity}}
0<\kappa_{s}(\mu)<\sqrt{4\pi},~~~~
0<\lambda_{s}(\mu)<\sqrt{4\pi},~~~~
\end{eqnarray}
for Renormalization Group (RG) scale $\mu$ between the EW and Plank mass scale.
We have used two-loop RG equations \cite{1203.5106},
\begin{eqnarray}{\label{RGE}}
\beta_{\kappa_{s}}&=&\frac{1}{(4\pi)^{2}}\left[-\left(\frac{9}{10}g_{1}^{2}+\frac{9}{2}g_{2}^{2}-6\lambda-6y^{2}_{t}-12\lambda_{s}\right)\kappa_{s}+4\kappa_{s}^{2}\right]
+\frac{1}{(4\pi)^{4}}\cdots \nonumber\\
\beta_{\lambda_{s}}&=&\frac{1}{(4\pi)^{2}}\left(\kappa_{s}^{2}+36\lambda_{s}^{2}\right)
+\frac{1}{(4\pi)^{4}}\cdots
\end{eqnarray}
where $g_i$ refer to the SM gauge coupling, and $y_t$ denotes the top Yukawa coupling.
As expected from Eq.(\ref{RGE}), 
the correlation between the upper bounds on $\kappa_s$ and $\lambda_s$ is mildly sensitive to $s$ scalar mass, but strongly to each other.
Our scan shows that the critical values are given by $\kappa_{s}<0.6$ and $\lambda_{s}<0.131$.

Of course, these bounds can be relaxed when the high energy scale above which the model is not effective is smaller than our reference value.

\section{Electroweak Vacuum Stability}
It was shown in \cite{higgscosm} that the EW vacuum is metastable for the observed Higgs mass $m_{h}\simeq 125$ GeV,
as seen from the RG equation for $\lambda$.
Since the sign of beta function $\beta_{\lambda}$ is always negative in the SM, 
$\lambda$ becomes negative above some critical RG scale $\mu_{*}$.
$\mu_{*}$ is approximately equal to the value $h_{*}$, at which the Higgs potential is maximal.
For the central values of top quark mass $m_{t}=173.2$ GeV 
and structure constant of QCD gauge coupling $\alpha=0.1184$, 
$h_{*}\simeq 10^{9}-10^{10}$ GeV at the two-loop level.
The situation can be improved to $h_{*} \simeq 10^{19}$ GeV 
for lighter $m_{t}$ within $1\sigma$ uncertainty \cite{1504.08093}.
So, in the most of SM parameter space EW vacuum is only metastable.

For EW vacuum being metastable, 
it suffers from large quantum fluctuation in the early Universe,
especially during and after inflation,
which leads to strong theoretic constraint on the model parameters.

Let us firstly discuss the survival probability of EW vacuum during inflation.
Denote $P(h,t)$ the probability for Higgs field with the value $h$ at time $t$,
which can be determined via solving the Fokker-Plank equation \cite{vacuum1, vacuum2, vacuum3},  
\begin{eqnarray}{\label{Fokker}}
\frac{\partial P}{\partial t}=\frac{\partial }{\partial h}\left[\frac{H_{c}^{3}}{8\pi^{2}}\frac{\partial P}{\partial h}+\frac{V' (h)}{3H_{c}}P\right].
\end{eqnarray}
Here $H_{c}$ refers to the Hubble parameter 
\footnote{The magnitude of Hubble constant is directly related to the scalar-to-tensor ratio $r$,  
with $H_{c}\simeq 1\times 10^{16} \times (r/0.01)^{1/4}$ GeV.
We take $H_{c}=10^{16}$ GeV as the reference point,
which is a good approximation unless $r$ is extremely small.},
which is approximate constant during inflation.
$V'$ denotes derivative of Higgs potential over Higgs field.
For the case $h< h_{*}$ the quantum fluctuation-$H^{3}_{c}$ term 
dominates the classical one-the last term in Eq.(\ref{Fokker}).
If one ignores this classical term,
Eq.(\ref{Fokker}) can be easily solved in terms of separating variants 
and resonable boundary conditions. 
See \cite{vacuum2, vacuum3} for details. 
Then, the probability for staying at the domain $h>h_{*}$ other than EW vacuum at the end of inflation is given by,
\begin{eqnarray}{\label{propability}}
P(\mid h\mid >h_{*})=1-\int^{h_{*}}_{-h_{*}}P (h, \frac{N}{H_{c}}) dh,
\end{eqnarray} 
where $N\simeq 55-60$ is the number of e-folds.

\begin{figure}[!h]
\centering
\includegraphics[width=4.5in]{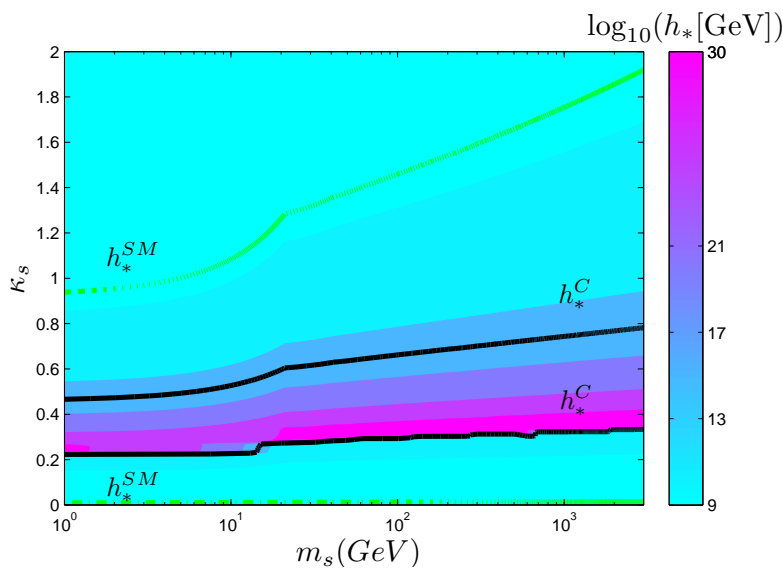}
 \caption{The value of $h_{*}$ at which the Higgs potential is maximal in the parameter space 
of $\kappa_{s}-m_{s}$. The green curves correspond to the case of SM. 
The black curves show the critical value as required by EW vacuum stability. 
It clearly shows that the value of $h_{*}$ can be uplifed into the region surrounded by the two black curves. }
 \label{figh}
\end{figure}

If the probability for staying at the metastable EW vacuum is not zero as required,
$P(\mid h\mid >h_{*})$ should be smaller than $e^{-3N}$, 
which gives rise to,
\begin{eqnarray}{\label{h}}
 h_{*} > \sqrt{\frac{3}{2}}\frac{N }{\pi}H_{c} \simeq 25 H_{c} \simeq 2.5\times 10^{17} \text{GeV}.
\end{eqnarray}
Obviously, this condition cannot be satisfied in the most of parameter space of SM.
The constraint in Eq.(\ref{h}) can be relaxed in the context of SM 
by introducing a new interation $\xi_{H}\mathcal{R}^{2}\mid H\mid ^{2}$  \cite{vacuum3, 0710.2484} between the Higgs and Ricci scalar $\mathcal{R}$.

With no need of parameter $\xi_{H}$,
this reduction can be alternatively realized in the HDM.
It is easy to understand in terms of the new contribution to beta function $\beta_{\lambda}$,
which reads at the two-loop level,
\begin{eqnarray}{\label{beta}}
\delta\beta_{\lambda}=\beta^{\text{(HDM)}}_{\lambda}-\beta^{\text{(SM)}}_{\lambda}=
\frac{\kappa^{2}_{s}}{(4\pi)^{2}}
+\frac{1}{(4\pi)^{4}}(-5\lambda\kappa^{2}_{s}-4\kappa^{3}_{s}).
\end{eqnarray}
The value of $h_{*}$ can be uplifted for small $\kappa_{s}$,
as the sign of the new contribution in Eq.(\ref{beta}) can be positive in this region.
In contrast, $h_{*}$ may be reduced in the region of large $\kappa_s$,
where the sign of the new contribution in Eq.(\ref{beta}) is negative.
The green curve in Fig.\ref{figh} clearly shows the critical value $\kappa_{s}^{c}\sim 0.9$
\footnote{For negative $\kappa_{s}$, $\mid\kappa_{s}^{c}\mid$ is larger 
in comparison with positive $\kappa_s$.
However, its magnitude is more seriously bounded above by the analysis of perturbativity.} .
Below this critical value $h_{*}$ can be larger than SM value $h_{*}^{\text{SM}}$.
More interesting, as shown by the red region corresponding to the range $\kappa_{s}\simeq 0.25-0.8$, 
the required value $h^{c}_{*}=2.5\times 10^{17}$ GeV in Eq.(\ref{h}) can be satisfied.

The effect on the vacuum stability of the singlet scalar potential  due to quantum fluctuations is more easily understood, in comparison with the Higgs potential.
Since $\left<s\right>=0$ is the true minimal of singlet scalar potential,
quantum fluctuation with magnitude of order $\delta s\sim H_c$ 
may push $s$ towards to an unstable vacuum from the origin.
But it rapidly returns to the origin vacuum through classical tunning process.
This understanding holds as long as $\lambda_{s}$ is always positive
between the EW and Plank scale.

Once inflation ends,  (p)reheating begins immediately. 
The quantum effect on the EW vacuum during (p) reheating can be studied in terms of the equations of motion for $s$ and Higgs in principle. 
Unfortunately, a concrete constraint on the model parameters in HDM can not be derived 
without knowledge of reheating temperature $T_\text{re}$,
which is an important physical quantity involved \cite{vacuum3}.
To determine $T_\text{re}$ 
interactions between the inflaton scalar and SM fields should be given explicitly.

\section{Dark Matter Phenomenology}
In this section we firstly consider the constraints arising from the DM relic density and 
direct detection limits at the LUX and Xenon100 experiments.
Then we discuss the prospect at the Xenon1T experiments.

\subsection{Relic Density}
The Plank and WMAP 9-year data have measured the DM relic density in high precision \cite{1303.5076},
\begin{eqnarray}{\label{relic}}
\Omega_{\text{DM}}h^{2}=0.1199\pm 0.0027
\end{eqnarray}
With the assumption that the relic density of $s$ scalar is totally produced by the thermal freeze-out process, we have
\begin{eqnarray}
\Omega_{s} h^{2}\sim 0.1~\text{pb}/\left<\sigma_{ann}\upsilon\right>，\nonumber
\end{eqnarray}
where $\sigma_{ann}$ is the total annihilation cross-section for $s$ and $\upsilon$ is the relative velocity.

\begin{figure}[!h]
\centering
\includegraphics[width=4.5in]{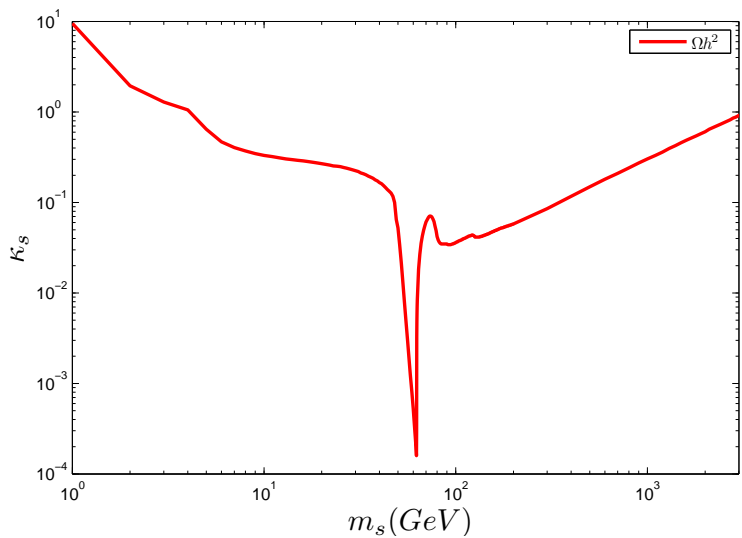}
 \caption{Contours of $s$ scalar relic density $\Omega_{s}h^{2}$ as required by the DM  projected to the two-parameter plane of $(m_{s}, \kappa_{s})$. }
 \label{figr}
\end{figure}

The $ss$ annihilation mainly proceeds through Higgs-mediated process in the $s$ channel.
Sub-dominant process include $s$ exchange in the $t$ channel 
and annihilation into $hh$.
So, $\sigma_{ann}$ is sensitive to the model parameters $\kappa_s$ and $m_s$.

Instead of analytic method we use the code MicrOMEGAs \cite{1407.6129} to calculate the $s$ scalar relic density $\Omega_{s}h^{2}$,
with the Feynman rules generated by the package LanHEP  3.2.0 \cite{1412.5016}.
In Fig.\ref{figr} we project the contour of relic density $\Omega_{s}h^{2}$ as required by Eq.(\ref{relic}) to the two-parameter plane of $(m_{s}, \kappa_{s})$. 
As expected the required value of $\kappa_{s}$ is minimal for $m_{s}$ near $m_{h}/2$,
because near this mass region the annihilation cross section is resonantly enhanced.
Plot similar to Fig.\ref{figr} is also shown, e.g., in \cite{1306.4710}
\footnote{
Although these plots match very well,
we would like to mention that contributions to $\sigma_{ann}$ due to three- and four-body final states or QCD corrections are not included in MicrOMEGAs,
with which the shape around $W$ boson mass is smoother \cite{1210.4196}.
We thank the referee for reminding us this point.
Since DM mass region near $80$ GeV has been excluded by the LUX experiment, these corrections are expected to not affect our main results in Sec.6.}.

Fig.\ref{figr} indicates that a wide mass range for $s$ scalar is viable.
However, it also shows that $\kappa_s$ bigger than $0.1$ is required 
when $m_s$ is far away from the {\it resonant mass region}. 
For example, large $\kappa_{s}\geq 0.5$ is needed for $m_{s}$ above $3$ TeV,
which is not favored neither by the EW vacuum stability nor the perturbative analysis.
In the next subsection, we discuss the direct detection in the allowed parameter space.

\begin{figure}[!h]
\centering
\includegraphics[width=4.5in]{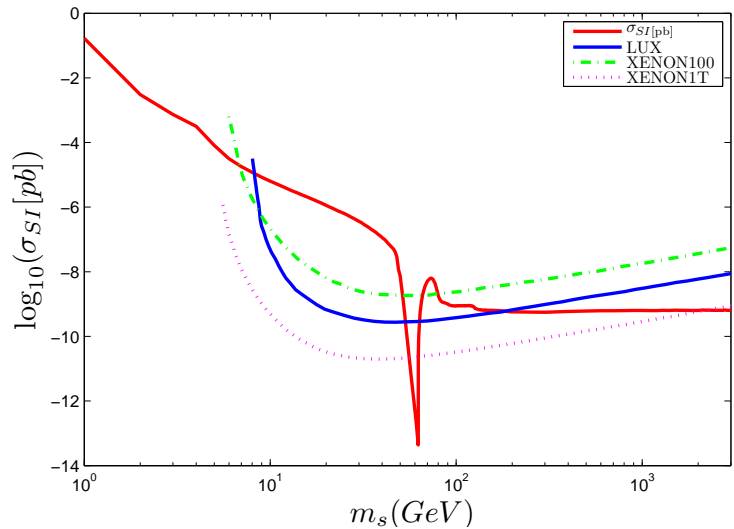}
 \caption{Spin-independent nucleon-DM scattering cross section as function of $m_{s}$. Various experimental limits are also shown for comparison.}
 \label{figSI}
\end{figure}

\subsection{Direct Detection}
The spin-independent nucleon-DM scattering cross section is given by,
\begin{eqnarray}{\label{crosssection}}
\sigma_{\text{SI}}=\frac{\kappa_{s}^{2}f^{2}_{N}\mu^{2}m^{2}_{N}}{4\pi m^{4}_{h}m^{2}_{s}},
\end{eqnarray}
where $m_{N}$ is the nucleon mass, 
$\mu=m_{s}m_{N}/(m_{s}+m_{N})$ is the DM-nucleon reduced mass,
and $f_{N}\sim 0.3$ \cite{1306.4710} is the hadron matrix element. 

The current best limit on $\sigma_{\text{SI}}$ comes from the Xenon100 \cite{Xenon100} and LUX \cite{LUX} experiments.
In Fig.\ref{figSI} we plot the spin-independent nucleon-DM scattering cross section as function of $m_{s}$, 
with various experimental limits for comparison. 
The red curve corresponds to the DM relic density, 
which indicates that there are three viable regions,
\begin{eqnarray}{\label{regions}}
\text{low~mass~region} &:& 1~\text{GeV}\leq m_{s} \leq 6~\text{GeV} , \nonumber\\
\text{resonant~mass~region} &:& 56~\text{GeV}\leq m_{s} \leq 66~\text{GeV} , \nonumber\\ 
\text{high~mass~region} &:& m_{s} \geq 185~\text{GeV}.
\end{eqnarray}
By combing the constraint from SM vacuum stability, one finds that 
\begin{itemize}
 \item The {\it  resonant mass region} is totally excluded,
which is more powerful than the indirect constraint from the Fermi-LAT \cite{1305.5597}.
 \item The {\it  high mass region} is further reduced to more narrow mass window,
which is the subject of Xenon 1T \cite{Xenon1T} experiment for $m_s$ below 4~TeV.
\item The {\it low mass region} can be consistent with the constraint from SM vacuum stability. However, we will show in Sec.V that the {\it low mass region} is excluded by the Higgs invisible decay $h\rightarrow ss$ at the LHC experiment (see also some literature in Refs.[8-18]).
\end{itemize}

\begin{figure}[!h]
\centering
\includegraphics[width=4.5in]{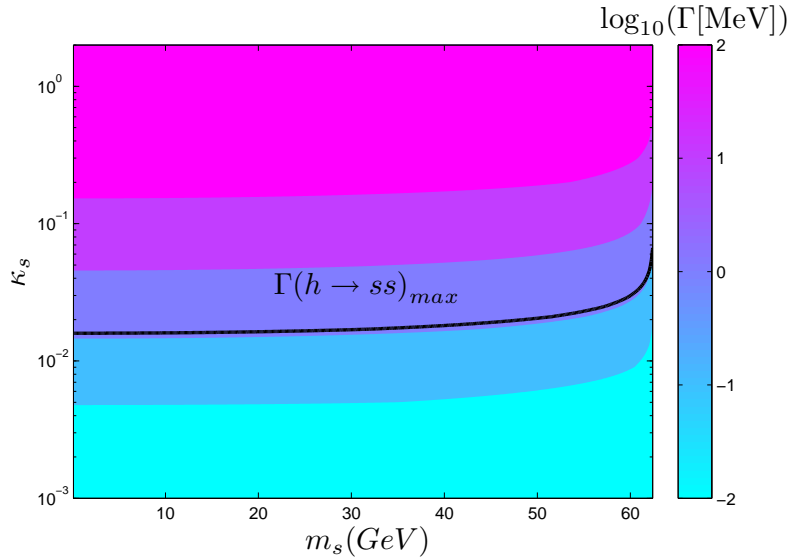}
 \caption{Contours of invisible decay width $\Gamma(h\rightarrow ss)$ in the two-parameter plane of $m_{s}-\kappa_{s}$. The contour in black denotes the latest upper bound in Eq.(\ref{invdecay}).
Region above this line is excluded. }
 \label{figd}
\end{figure}

\section{LHC Phenomenology}
Since the scalar singlet only couples to the Higgs directly,
the signal searches for this scalar at the LHC 
mainly focus on the Higgs invisible decay $h\rightarrow ss$ 
in the {\it low mass region}  and {\it resonant mass region} with $m_{s}<m_{h}/2$. 
Note that for scalar singlet DM there is no mixing effect
\footnote{If $s$ scalar is not DM,
one may discuss its phenomenology by introducing mixing effects. 
For a recent discussion, see \cite{Zheng:2015}.}
between $s$ and $h$, 
as clearly shown in Eq.(\ref{Lagrangian1})  and Eq.(\ref{Lagrangian2}).

Very recently, the ATLAS Collaboration has reported the latest data about the upper bound 
on the Higgs invisible decay width \cite{invdecay},
\begin{eqnarray}{\label{invdecay}}
\Gamma_{\text{inv}}\leq 0.29~ \Gamma_{\text{SM}},
\end{eqnarray}
which is obviously stronger than what was used in some earlier literature. 
An update is thus meaningful.
In Eq.(\ref{invdecay}) $\Gamma_{\text{SM}}\simeq 4.15$ MeV for 125 GeV Higgs mass,
and the invisible decay width  $\Gamma_{\text{inv}}$  is given by,
\begin{eqnarray}{\label{width}}
\Gamma(h\rightarrow ss)=\frac{\kappa_{s}^{2}\upsilon^{2}}{32\pi m_{h}}\sqrt{1-\frac{4m^{2}_{s}}{m^{2}_{h}}}.
\end{eqnarray}

We show the decay width $\Gamma(h\rightarrow ss)$ in Fig.\ref{figd}.
The contour in black denotes the latest upper bound in Eq.(\ref{invdecay}),
and region above it is excluded. 
In particular, this bound excludes
the whole {\it low mass region}  and the {\it resonant mass region} with $m_{s}\leq 62.5$ GeV in Eq.(\ref{regions}).
In contrast, the {\it high mass region} is less constrained at the LHC in compared with the {\it low mass region}.

\begin{figure}[!h]
\centering
\includegraphics[width=4.5in]{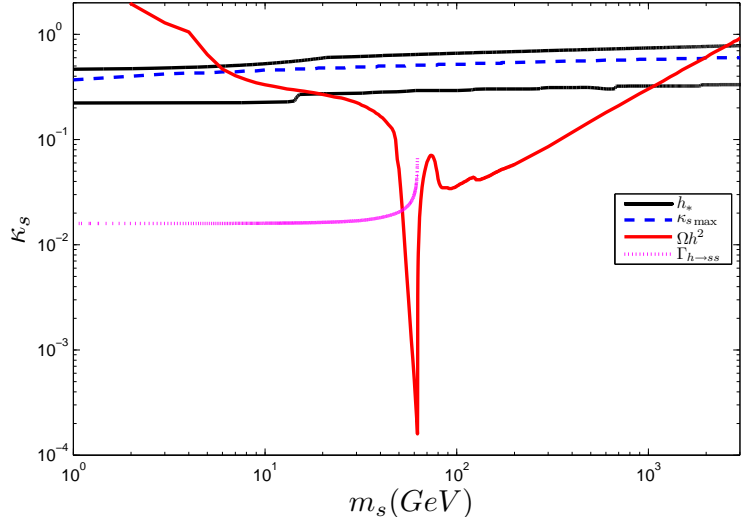}
 \caption{Mass regions consistent with various constraints. 
In the light of the SM vacuum stability,  
only mass region $1.1$ TeV $\leq m_{s}\leq$ $2.5$ TeV is viable, 
and it slightly reduces to $1.1$ TeV $\leq m_{s}\leq$ $2.0$ TeV by the additional requirement of perturbativity. 
For the explanation about $h_{*}$ and $\kappa_{s \text{max}}$, 
see previous discussions.}
 \label{figtot}
\end{figure}

\section{Conclusions and Discussions}
In this paper we have reconsidered the simplest HDM.
By combining the constraints arising from DM relic density,  
direct detection limit on spin-independent nucleon-DM scattering cross section from the LUX,
and limit on the Higgs invisible decay from the LHC,
we have obtained two viable mass regions (see Fig.\ref{figtot}):
\begin{eqnarray}{\label{r1}}
\text{resonant~mass~region} &:& 62.5~\text{GeV}\leq m_{s} \leq 66~\text{GeV} , \nonumber\\ 
\text{high~mass~region} &:& m_{s} \geq 185~\text{GeV}.
\end{eqnarray}

We also show that if the constraint from the SM vacuum stability is imposed,
we arrive at the main conclusion- 
the {\it resonant mass region} is totally excluded 
and the {\it high mass region} in Eq.(\ref{r1}) is reduced to a narrow window,
\begin{eqnarray}{\label{r2}}
\text{high~mass~region} &:& 1.1~\text{TeV}\leq m_{s} \leq 2.55~\text{TeV} ,
\end{eqnarray}
as shown in Fig.\ref{figtot}.
If one further imposes the constraint due to perturbativity,
the mass range in Eq.(\ref{r2}) is slightly reduced to,
\begin{eqnarray}{\label{r3}}
\text{high~mass~region} &:& 1.1~\text{TeV}\leq m_{s} \leq 2.0~\text{TeV}.
\end{eqnarray}

Fortunately, as shown in Fig.\ref{figSI} the mass region in Eq.(\ref{r3}) can be totally detected by the Xenon 1T experiment in the near further.
Similarly to Xenon 1T experiment, this {\it high mass region} can be also directly detected at colliders in principle.
As a result of small production cross section of order less than $1$ fb for the $s$ scalar
at the 14-TeV LHC \cite{1112.3299},
a $5\sigma$ discovery acquires luminosity at least of order $\mathcal{O}(10)$ $ab^{-1}$.

Independently, effects on large scale structure induced by DM self-interaction is a new window for the detection.
In compared with the abelian vector- or fermion-like DM,
$s$ scalar with quartic self-interaction is very distinctive. 
DM self-interaction has been used to interpret \cite{1504.06576} the observations on four galaxies in the core of galaxy cluster Abell 3827 \cite{1504.03388}.
If the new observations indeed arise from the DM self-interaction 
other than astrophysical artifact,
the {\it high mass region} is highly constrained, 
see, e.g., for recent discussion \cite{1505.01793}. 

In variety of contexts the {\it high mass region} in Eq.(\ref{r2}) may be modified.
For example, it can be relaxed 
if one employs additional mechanism to stabilize the SM vacuum by introducing more new parameters, such as the non-minimal interaction between the Higgs and Ricci scalar with  non-negligible coupling constant.
In contrast, this  {\it high mass region} may be alternatively strengthened 
if one assumes \cite{1306.4710} that the DM relic abundance is partially  other than totally saturated by the scalar singlet. 
In this case scalar $s$ may play a positive role in the electroweak baryogenesis \cite{1210.4196}.
\\

~~~~~~~~~~~~~~~~~~~~~~~~~~~~~~~~~~~~
$\mathbf{Acknowledgments}$\\
HH would like to thank Felix Kahlhoefer for suggesting the use of code LanHEP. 
This work is supported in part by the Natural Science Foundation of China under grant No.11247031 and 11405015.

\end{document}